\documentclass[aps,prb,amsmath,twocolumn,amssymb,floatfix,showpacs,superscriptaddress,nofootinbib,longbibliography,10pt]{revtex4-2}
\usepackage{MnSymbol}
\usepackage{braket}
\usepackage[dvipsnames]{xcolor}
\usepackage{float}
\usepackage{tikz}
\usepackage[colorlinks=true,linktoc=page,linkcolor=blue,citecolor=blue,urlcolor=blue]{hyperref}
\usepackage{multirow}
\usepackage{subfigure}
\usepackage{lipsum}
\usepackage{orcidlink}
\usepackage[normalem]{ulem}
\usepackage{amsmath} 
\usepackage{setspace}
\newcommand{\vect}[1]{\boldsymbol{\mathrm{#1}}}

\mathchardef\mhyphen="2D 

\newcommand{\ie}{{i.e.,\,\,}}

\newcommand\bea{\begin{eqnarray}}
\newcommand\eea{\end{eqnarray}}
\newcommand\beq{\begin{equation}}  
\newcommand\eeq{\end{equation}}

\newcommand{\non}{\nonumber}  

\definecolor{lime}{HTML}{A6CE39}
\usepackage{sidecap,tikz}
\DeclareRobustCommand{\orcidicon}{\hspace{-1.0mm}
    \begin{tikzpicture}
        \draw[lime, fill=lime] (0.0,0.0) 
        circle [radius=0.15] 
        node[white] {{\fontfamily{qag}\selectfont \tiny \,ID}};
        \draw[white, fill=white] (-0.0525,0.095) 
        circle [radius=0.007];
    \end{tikzpicture}
    \hspace{-3.0mm}
}
\foreach \x in {A, ..., Z}{\expandafter\xdef\csname orcid\x\endcsname{\noexpand\href{https://orcid.org/\csname orcidauthor\x\endcsname}{\noexpand\orcidicon}}
}

\allowdisplaybreaks

\begin{document}

\title{Dissipation induced Majarona $0$- and $\pi$-modes in a driven Rashba nanowire}  
\author{Koustabh Gogoi~\orcidlink{0009-0002-4590-5607}}
\affiliation{Department of Physics, BITS Pilani-Hyderabad Campus, Telangana 500078, India}

\author{Tanay Nag~\orcidlink{0000-0001-6052-7232}}
\email[Corresponding author: ]{tanay.nag@hyderabad.bits-pilani.ac.in}
\affiliation{Department of Physics, BITS Pilani-Hyderabad Campus, Telangana 500078, India}

\author{Arnob Kumar Ghosh~\orcidlink{0000-0003-0990-8341}}
\email[Corresponding author: ]{arnob.ghosh@physics.uu.se}
\affiliation{Department of Physics and Astronomy, Uppsala University, Box 516, 75120 Uppsala, Sweden}

\begin{abstract}
Periodic drive is an intriguing way of creating topological phases in a non-topological setup. However, most systems are often studied as a closed system, despite being always in contact with the environment, which induces dissipation. Here, we investigate a periodically driven Rashba nanowire in proximity to an $s$-wave superconductor in a dissipative background. The system's dynamics is governed by a periodic Liouvillian operator, from which we construct the Liouvillian time-evolution operator and use the third-quantization method to obtain the `Floquet damping matrix', which captures the spectral and topological properties of the system. We show that the system exhibits edge-localized topological Majorana $0$-modes~(MZMs) and $\pi$-modes~(MPMs). Additionally, the system also supports a trivial $0$-modes~(TZMs) and $\pi$-modes~(TPMs), which are also localized at the edges of the system. The MZMs and the MPMs are connected to the bulk topology and carry a bulk topological invariant, while the emergence of TZMs and TPMs is primarily tied to exceptional points and is topologically trivial. We show that both the topological (MZMs and MPMs) and trivial (TZMs and TPMs) edge modes are robust against onsite disorder. We study the topological phase diagrams in terms of the topological invariants and show that the dissipation can modify the topological phase diagram substantially and even induce topological phases in the system. Our work extends the understanding of a driven-dissipative topological superconductor.
\end{abstract}

\maketitle

\section{Introduction} \label{Sec:introduction}
The search for Majorana zero-modes~(MZM) in topological superconductors~(TSCs) has engaged physicists for the past two decades~\cite{Kitaev_2001,qi2011topological,Leijnse_2012,Alicea_2012,beenakker2013search,ramonaquado2017,tanaka2024theory}. The seminal proposition for realizing Majorana in a condensed matter system, proposed first by Kitaev, is based on $p$-wave superconductor~\cite{Kitaev_2001}. However, the $p$-wave superconductors do not emerge naturally. Nevertheless, we can still engineer an effective $p$-wave superconductor from a conventional $s$-wave superconductor by employing a semiconducting nanowire with Rashba spin-orbit coupling~(SOC) and a Zeeman field~\cite{Oreg2010,LutchynPRL2010,Leijnse_2012,Alicea_2012,Mourik2012Science,das2012zero,ramonaquado2017}. Although, the experimental evidence of detecting MZM in this heterostructure setup remain elusive~\cite{Mourik2012Science,das2012zero,DengNano2012,Rokhinson2012,Finck2013,Albrecht2016,Deng2016,NichelePRL2017,JunSciAdv2017,Zhang2017NatCommun,Gul2018,Grivnin2019,ChenPRL2019}. Nonetheless, a simpler and more controllable setup to realize so-called ``poor man's'' MZMs without any topological protection and based on the Kitaev model has attracted a lot of attention recently~\cite{LeijnsePMPRB2012,DvirNature2023}.

At the same time, the quest for TSC has been further enriched in periodically driven systems, allowing us to engineer a non-trivial phase in an otherwise trivial system~\cite{JiangPRL2011,ThakurathiPRB2013,LiuPRL2013,benito14,PotterPRX2016,ThakurathiPRB2017,YangPRL2021,GhoshDynamical2022,MondalPRB2023,MondalPRB2023-2,GhoshTimedynamics2023,RoyPRB2024}. Moreover, the driven system also offers topologically protected Majorana states at a finite quasienergy $E=\omega/2$ ($\omega$ is the frequency of the drive), which are called Majorana $\pi$-modes~(MPMs)~\cite{YangPRL2021,GhoshDynamical2022,MondalPRB2023,MondalPRB2023-2,GhoshTimedynamics2023,AhmedPRB2025}. These MPMs with the concurrent MZMs offer more flexibility in topological quantum computations~\cite{BomantaraQC2018,BauerBelaPRB2019}. The MZMs and MPMs in a driven system have also been explored in the presence of non-Hermitian~(NH) effects~\cite{ZhouPRB2020,Ghosh2022NHPRBL}. The NH system offers many exotic behaviors that are absent in their Hermitian counterparts, such as the breakdown of conventional bulk boundary correspondence~\cite{YaoSPRL2018,KunstPRL2018,BergholtzRMP2021,OkumaAnnualRev2023}, NH skin effects~\cite{LeePRB2019,LiNatCommun2020,BergholtzRMP2021,OkumaAnnualRev2023,BanerjeeJPCM2023}, the emergence of exceptional points~(EPs)~\cite{TonyLeePRL2016,Hodaei2017,Ghatak2019,AshidaYotoAP2020,DennerNatComm2021,BergholtzRMP2021,SayyadPRR2022,OkumaAnnualRev2023,BanerjeeJPCM2023}, and non-trivial behavior in dynamical quantum phase transition~\cite{MondalNHPRB2022,MondalPTPRB2023}. Moreover, the NH-terms can substantially modify the topological phases in both driven and undriven systems~\cite{YaoSPRL2018,Ghosh2022NHPRBL,RoyPRB2025}.

The NH effect can enter a system in many different ways, but the dissipation via the coupling with the environment offers an intriguing way of obtaining the NH effect in a system~\cite{SanJoseSP2016,DangelPRA2018,MoosSciPost2019,KawasakiPRB2022,YangPRR2023,HegdePRL2023}.
While the dissipation can have a detrimental effect on the system and can manipulate the system's topological properties~\cite{SongPRL2019,LieuPRL2020,MingantiPRA2019,LiuChunPRR2020,HagaPRL2021,OkumaPRB2021,ZhouPRA2022,YangPRR2022,LeePRB2023,KawabataPRX2023,YangFeiPRB2023}, specifically engineered dissipation has also been employed to induce a topological phase in an otherwise trivial system~\cite{Diehl2008,KrausPRA2008,Verstraete2009,Weimer2010,Diehl2011,BardynPRL2012,BardynNJP2013,BudichPRA2015,IeminiPRB2016,GhoshDissipation2024}.
In a dissipative system, the system's long-time evolution is governed by the density matrix, and in the Markovian approximation of a memoryless bath~\cite{breuer2002theory,KrausPRA2008}, the density matrix is governed by the Liouvillian operator, which follows the Lindblad master equation~\cite{Lindblad1976,Gorini1976}. Moreover, by employing the third-quantization method for linear jump operators, we can obtain a matrix form of the Liouvillian operator, which behaves like the NH Hamiltonian~\cite{ProsenNJP2008,ProsenJSM2010,ProsenNJP2010}. A topological classification based on the Liouvillian has also been investigated~\cite{LieuPRL2020}.

In this context, the MZMs and MPMs in a driven system are usually studied as a closed system. However, the system is often in contact with the surroundings, introducing some amount of dissipation~\cite{breuer2002theory,LandiRMP2022}. On the other hand, the dissipation offers a way of engineering the NH effect in a driven system~\cite{MoosSciPost2019,GhoshDissipation2024}. Recently, an open Rashba nanowire coupled with the environment has been studied in the presence of dissipation, while it has also been demonstrated that the dissipation can lead to the emergence of MZMs in such a system~\cite{GhoshDissipation2024}. Motivated by this, we ask the following interesting questions: (a) Can the MZMs and MPMs be obtained in a driven-dissipative Rashba nanowire? (b) Can the dissipation even assist in obtaining topological phases that are not present in the non-dissipative case? In this manuscript, we affirmatively answer these questions.

In this work, we consider a three-step driving protocol in a Rashba nanowire, which is coupled with the environment via linear jump operators. We employ the third-quantization method and obtain a Floquet Liouvillian~\cite{ProsenNJP2008,ProsenNJP2010,ProsenJSM2010,MoosSciPost2019}. We demonstrate that the dissipation offers an exotic way of obtaining and understanding the NH effect in a driven system where MZMs and MPMs can be engineered.  Surprisingly, we also find some boundary modes appearing at both $0$- and $\pi$-quasienergy, dubbed as trivial $0$-modes~(TZMs) and $\pi$-modes~(TPMs)  respectively, that are not protected by the bulk gap but rather induced by the EPs.  We also investigate the eigenvalue spectra and the edge-localization properties of the states appearing at quasienergies $0,\pm \pi$. While the TZMs and TPMs do not carry any topological indices, we topologically characterize the MZMs and MPMs employing winding numbers based on a generalized chiral symmetry and obtain a phase diagram in the parameter space. This phase diagram also points out the non-trivial generation of MZMs and MPMs in a dissipative system.

The remainder of the manuscript is organized as follows. In Sec.~\ref{Sec:Model}, we discuss the model Hamiltonian for the Rashba nanowire and introduce the driving protocol. In Sec.~\ref {Sec:FloquetLindbladian} and Sec.~\ref{Subsec:TopoChar}, we discuss the form of dissipation, the third-quantization method to obtain the Floquet Liouvillian, and the real-space topological invariant to characterize the system. Section~\ref{Sec:Results} presents the main results of this work. In Sec.~\ref{Sec:Exp}, we discuss the plausible experimental implementation of our work and the feasible, realistic parameters. We conclude with a summary and discussion in Sec.~\ref{Sec:Summary}.

\section{Model and driving protcol} \label{Sec:Model}
We consider a Rashba nanowire in proximity to a bulk $s$-wave superconductor in the presence of an in-plane magnetic field. The Hamiltonian of this system reads as~\cite{Oreg2010,LutchynPRL2010,Leijnse_2012,Alicea_2012,Mourik2012Science,das2012zero,ramonaquado2017} 
\begin{align}
	H_0=& -\mu \sum_{i=1}^{N} \sum_{\alpha=\uparrow,\downarrow} c^\dagger_{i \alpha}  c_{i \alpha} + t_{h} \sum_{i=1}^{N-1} \sum_{\alpha=\uparrow,\downarrow} c^\dagger_{i \alpha}  c_{i+1 \alpha}  \non \\
    &-i \lambda_{R} \sum_{i=1}^{N-1} \sum_{\alpha,\beta=\uparrow,\downarrow} c^\dagger_{i \alpha } (\sigma_z)_{\alpha, \beta}  c_{i+1 \beta}  \non \\
	& +B  \sum_{i=1}^{N} \sum_{\alpha,\beta=\uparrow,\downarrow} c^\dagger_{i \alpha} (\sigma_x)_{\alpha, \beta} c_{i \beta}  + \Delta \sum_{i=1}^{N} c^\dagger_{i \uparrow} c^\dagger_{i \downarrow} + {\rm H.c.} ,
 \label{Eq:Rashbarealspace}
\end{align}
where $\mu$, $t_h$, $\lambda_R$, $B$, and $\Delta$ represent onsite chemical potential, hopping amplitude, Rashba SOC strength, in-plane magnetic field along $x$-direction, and proximity induced $s$-wave pairing gap, respectively. Here, $c_{i \alpha}$~($c_{i \alpha}^\dagger$) creates~(annihilates) an electron at lattice site $i$ with spin $\alpha=\uparrow,\downarrow$, and $N$ denotes the number of lattice sites in the system. The Pauli matrices $\vect{\sigma}$ act on the spin space. For an infinite and translation-invariant nanowire, we can write the Hamiltonian of the system in the momentum space considering the Bogoliubov–de Gennes basis $\bf{\Psi_{k}}$ = $\left\{ c_{k\uparrow}, c_{k\downarrow}, -c^{\dagger}_{-k\downarrow}, c^{\dagger}_{-k\uparrow} \right\}^T$ as 
\begin{align}
    H_0(k)=& (-\mu + 2t_{h}\cos k) \ \tau_{z}\sigma_0 + 2 \lambda_{R} \sin k \  \tau_{z} \sigma_{z} \non \\
    &+ B \ \tau_{0}\sigma_{x} +\Delta \ \tau_{x}\sigma_{0} ,
    \label{hamiltonian}
\end{align} 
where the Pauli matrices $\vect{\tau}$ act on the particle-hole sector. The Hamiltonian [Eq.~\eqref{hamiltonian}] respects chiral symmetry $S H_0(k) S = -H_0(k)$ with $S=\tau_y \sigma_z$ and particle-hole symmetry $C^{-1} H_0(k) C = -H(-k)$ with $C=\tau_y \sigma_y K$; $K$ being the complex-conjugation operator. However, the introduction of a magnetic field breaks the time-reversal symmetry $\mathcal{T}^{-1} H(k) \mathcal{T} \neq H(-k)$ with $\mathcal{T}=i \tau_0 \sigma_y K$. The interplay of Rashba SOC, magnetic field, and the $s$-wave superconductivity induces an effective spin-triplet $p$-wave pairing in the nanowire~\cite{Oreg2010,LutchynPRL2010,Leijnse_2012,Alicea_2012,Mourik2012Science,das2012zero,ramonaquado2017}. The system becomes topological supporting non-trivial MZMs when $\left| B_{c1} \right| < B < \left| B_{c2} \right|$ with $\left| B_{c1} \right| = \sqrt{(\mu - 2 t_h)^2 + \Delta^2}$ and $\left| B_{c2} \right| = \sqrt{(\mu + 2 t_h)^2 + \Delta^2}$~\cite{MondalPRB2023}. We set $2 t_h = \lambda_{R} = \Delta=1$ and $\mu=0$ throughout the manuscript unless mentioned otherwise. However, our results are independent of such a choice of the parameters.

To engineer the MZM and MPM in the Rashba nanowire, we consider a periodic drive in the form of a three-step drive protocol on top static Hamiltonian $H_0$ as~\cite{MondalPRB2023}
\begin{align}
    H(t) =& H_1  &t \in  \left[ 0, T/4 \right), \non \\
         =& H_0  &t \in  \left[ T/4, 3T/4 \right) , \non \\
         =& H_1 &t \in   \left[ 3T/4 , T \right] ,
    \label{Eq:Drive}
\end{align}
where the step Hamiltonians at the first and last step only comprise the onsite chemical potential, such that $H_1=-\mu_1 \sum_{i=1}^{N} \sum_{\alpha=\uparrow,\downarrow} c^\dagger_{i \alpha}  c_{i \alpha}$. In the absence of any loss in the setup, the spectral properties of the driven system is captured by the stroboscopic Floquet operator in the time-ordered~(TO) notation as $U(T,0)= {\rm TO} ~ \exp \left[ -i \int_{0}^{T} dt \  H(t) \right]$~\cite{MondalPRB2023,Ghosh_2024}. The Floquet operator follows the eigenvalue equation: $U(T,0) \ket{\psi_n} = \exp(-i \epsilon_n T) \ket{\psi_n}$, where $\epsilon_n$ is the quasienergy corresponding to the state $\ket{\psi_n}$. This system has been investigated in detail in Ref.~\cite{MondalPRB2023} and has been shown to support both MZMs and MPMs.

\section{Dissipation and Floquet Lindbladian} \label{Sec:FloquetLindbladian}
To explore the effect of dissipation on the MZMs and MPMs, we couple our system to an environment. However, to study the driven system in a dissipative background, we need to formulate the Floquet formalism in such a way that it captures the effect of the environment. In this section, we briefly discuss how to obtain the Floquet Liouvillian, which governs the dynamics of the system in the presence of dissipation~\cite{MoosSciPost2019,SchnellPRB2020}, employing the third-quantization method~\cite{ProsenNJP2008,ProsenJSM2010,ProsenNJP2010}. 

\subsection{Lindblad master equation and jump operator}
We begin by discussing the effect of the dissipation on a static system.  We adopt the Markovian approximation such that the time evolution of the density matrix $\rho(t)$ is governed by the Lindblad master equation~\cite{breuer2002theory,KrausPRA2008,Lindblad1976,Gorini1976}
\begin{subequations}
    \begin{align}
	\frac{\partial \rho (t)}{\partial t} \! & = \! -i \left[ H, \rho (t) \right] \! - \! \frac{1}{2} \!\sum_m \!\left( \left\{ L^\dagger_m L_m , \rho(t) \right\}\!-\!2 L^\dagger_m \rho(t) L_m \right) , \label{Lindblad1} \\
	& \equiv \mathcal{L} \rho (t)  \label{Lindblad2} ,
\end{align}
\end{subequations}
The first term in Eq.~\eqref{Lindblad1} involving the Hamiltonian $H$ generates a unitary evolution, while the second term associated with the Lindblad jump operator $L_m$ accounts for the system's coupling to the reservoir. More precisely, the term $\left\{ L^\dagger_m L_m , \rho(t) \right\}$ is partially responsible for a non-unitary evolution of the system, while the term $L^\dagger_m \rho(t) L_m $ is called the quantum jump~\cite{MingantiPRA2019}, which mixes between different states of the system. Together, these operators form the Liouville operator $\mathcal{L}$, which governs the dynamics of the system.

The Lindblad jump operator can have different forms, but for simplicity, we consider an onsite loss operator in terms of a linear combination of fermionic annihilation operators only, which at a lattice site $i$ reads as
\begin{align}
    L_{i}= \sqrt{\gamma} \left(c_{i\uparrow}+c_{i\downarrow}  \right), 
    \label{lossOperator}
\end{align}
where $\gamma$ denotes the strength of the dissipation. The loss operator $L_i$ acts on both spin species~\cite{GhoshDissipation2024}. The linear jump operators in Eq.~\eqref{lossOperator} allow us to obtain a matrix form of the Liouvillian $\mathcal{L}$, utilizing the third-quantization method~\cite{ProsenNJP2008,ProsenJSM2010,ProsenNJP2010}. 

\subsection{Third quantization}
Here, we briefly discuss the third-quantization method~\cite{ProsenNJP2008,ProsenJSM2010,ProsenNJP2010}. First, we replace the complex fermionic operator with real Majorana fermions as $c_{i\uparrow} = \left( w_{iA}+ iw_{iB} \right) /2$ and $c_{i\downarrow} = \left( w_{iC}+ iw_{iD} \right) /2$ such that the Majorana operators satisfy the anti-commutation rule: $\{w_{i\alpha}, w_{j\beta}\} = 2 \delta_{i,j}\delta_{\alpha,\beta}$; with $\alpha = A, B, C, D$ and $i,j=1,2, \cdots, N$. The Hamiltonian $H$ and the Lindblad jump operators $L_m$ can be written by employing the Majorana operators as $ H_M = \sum_{a,b}^{4N}w_{a}H_{a,b}w_{b}=\underline{w}^{T}{H}^{(w)}\underline{w}$ and $L = \sum_{a}^{4N} l_{a}w_{a} = l \underline{w}$; with  $\underline{w}=\left\{w_{1A},w_{1B},w_{1C},w_{1D},\cdots,w_{iA},w_{iB},w_{iC},w_{iD},\cdots\right\}^{T}$. Next, we introduce a Fock space called the Liouville-Fock space $\mathcal{K}$ spanned by the polynomial: $P_{\underline{\alpha}} = \prod_{a=1}^{4N} w_{a}^{\alpha_a}$, with $\alpha_{a}\in {0,1}$ representing the absence or presence of the Majorana operator $w_{a}$. The density matrix $\rho(t)$ is a vector in this Fock space $\mathcal{K}$. Within this Fock space, we can define the adjoint fermion creation operator: $\hat{\phi}_{a}|P_{\underline{{\alpha}}}\rangle = \delta_{\alpha_{a},0}|w_{a}P_{\underline{\alpha}}\rangle$ and annihilation operator: $\hat{\phi}_{a}^{\dagger}|P_{\underline{{\alpha}}}\rangle = \delta_{\alpha_{a},1}|w_{a}P_{\underline{\alpha}} \rangle$, such that $\{\hat{\phi}_{a},\hat{\phi}_{b}^{\dagger}\}=\delta_{a,b}$. Now, assuming an even number of Majorana fermions in the system, such that we can recast them to obtain complex fermions without leaving any unpaired Majorana fermions. We provide further details in Appendix~\ref{App:LiouvillianApp}. In the even subspace $(+)$, the Liouvillian operator, denoted as $\mathcal{L}_+$ can be put in the form~\cite{ProsenNJP2008,ProsenJSM2010,ProsenNJP2010,GhoshDissipation2024}
\begin{align}
	\mathcal{L}_+= \frac{1} {2}  
	\begin{pmatrix} 
		\underline{\phi}^\dagger  &  \underline{\phi}
	\end{pmatrix} 
	\begin{pmatrix} 
		-iX^\dagger &   Y \\
		0 & iX
	\end{pmatrix} 
	\begin{pmatrix} 
		\underline{\phi}^\dagger \\  \underline{\phi}
	\end{pmatrix}  -A_0 ,
 \label{Liouvillianplus}
\end{align} 
where $X=4 H_{M} +i (M+M^T)$, $Y=2 (M-M^T)$, and $A_0=\frac{1}{2} {\rm Tr} [X]$, with $M_{mn}=(l)_{m}^{T}(l)^{*}_{n}$. The vectors $\underline{\phi}^\dagger_k$ and  $\underline{\phi}_k$ contains $\hat{\phi}^\dagger_{a}$'s and $\hat{\phi}_{a}$'s, respectively. The operator $X$ is also called the damping matrix. Owing to the upper triangular form of $\mathcal{L}_+$, the spectral and topological properties of the system are solely determined by $X$~\cite{LieuPRL2020,GhoshDissipation2024}. Furthermore, the operator also resembles the NH Hamiltonians, but with a subtle difference that the imaginary eigenvalue of the $X$ is always positive, such that the density matrix decays over time and reaches the steady state: $\rho_{\rm SS}=\exp[i \mathcal{L}_+ t] \rho(0)$.

\subsection{Floquet-Liouvillian}
Here, we discuss how we obtain the Floquet-Liouvillian operator, which governs the dynamics of a driven-dissipative system. In this regard, we consider that the driven Rashba nanowire setup is constantly in contact with the environment such that the dissipation of the form given in Eq.~\eqref{lossOperator} is present in all the step Hamiltonians of Eq.~\eqref{Eq:Drive}. The description of the system involves a time-dependent periodic Liouvillian operator $\hat{\mathcal{L}}(t)$, and the  Liouvillian time-evolution operator $U_{\mathcal{L}}(T, 0)$ for this system can be constructed as~\cite{MoosSciPost2019}
\begin{align}
U_{\mathcal{L}}(T, 0) = \mathrm{TO} \exp \left[ \int_{0}^{ T} dt \, \hat{\mathcal{L}}(t) \right] .
\end{align} 
Here, $U_{\mathcal{L}}(T, 0)$ provides the stroboscopic description of the system in this driven-dissipative setup. We can obtain the Floquet Liouvilian $\mathcal{L}_F$ as: $U_{\mathcal{L}}(T,0)=e^{\mathcal{L}_F T}$. Now, considering the third-quantized Liouvillian operator [see Eq.~\eqref{Liouvillianplus}], we can obtain the Floquet Liouvillian as
\begin{align}
U_{\mathcal{L}}( T, 0)=e^{\mathcal{L}_FT} = 
\begin{pmatrix}
e^{-i X_F^\dagger T} & f(X, Y, T) \\
0 & e^{i X_F T}
\end{pmatrix}.
\end{align}
The Liouvillian time-evolution operator $U_{\mathcal{L}}( T, 0)$ being an upper triangular matrix, the spectral and topological properties of the system are governed by the `Floquet damping matrix' $X_F$ alone, and henceforth we focus our discussion on $X_F$. The Floquet damping matrix $X_F$ is the dynamical analog of the static damping matrix $X$ defined in Eq.~\eqref{Liouvillianplus}. The function $f(X,Y,T)$ is a complex function of $X$ , $Y$ and $T$; however, its explicit form is not required for the purposes of this work.

Having discussed the formalism, we now explicitly show the form of the Floquet damping matrix $X_F$ for our system. As discussed earlier, we allow the system to have dissipation in all the step Hamiltonians in Eqs.~\eqref{Eq:Drive}. Thus, the corresponding step damping matrices, written in the Majorana basis, owing to the dissipation of the form Eq.~\eqref{lossOperator}, read as
\begin{widetext}
    \begin{subequations}
\begin{align}
    X_0=& \mu \sum_{i=1}^{N} \sum_{\alpha \beta=1}^{4} w_{i \alpha} \left(\tilde{\tau}_{0} \tilde{\sigma}_{y} \right)_{\alpha,\beta}  w_{i \beta} - t_{h} \sum_{i=1}^{N-1} \sum_{\alpha, \beta=1}^{4} w_{i \alpha}   \left(\tilde{\tau}_{0} \tilde{\sigma}_{y} \right)_{\alpha,\beta}  w_{i+1 \beta}  -i \lambda_{R} \sum_{i=1}^{N-1} \sum_{\alpha \beta=1}^{4} w_{i \alpha}   \left(\tilde{\tau}_{z} \tilde{\sigma}_{0} \right)_{\alpha,\beta}  w_{i+1 \beta}  \non \\
    & -B \sum_{i=1}^{N} \sum_{\alpha, \beta=1}^{4} w_{i \alpha} \left(\tilde{\tau}_{x} \tilde{\sigma}_{y} \right)_{\alpha,\beta} w_{i \beta} +\Delta\sum_{i=1}^{N} \sum_{\alpha, \beta=1}^{4} w_{i \alpha} \left(\tilde{\tau}_{y}\tilde{\sigma}_{x} \right)_{\alpha,\beta} w_{i \beta}+\frac{i\gamma}{2} \sum_{i=1}^{N} \sum_{\alpha, \beta=1}^{4} w_{i \alpha} \left(\tilde{\tau}_{0} \tilde{\sigma}_{0} + \tilde{\tau}_{x} \tilde{\sigma}_{0} \right)_{\alpha,\beta} w_{i \beta}+ {\rm H.c.} ,  \label{SubEq:DampingmatrixStep1} \\
    X_{1} =& \mu_1 \sum_{i=1}^{N} \sum_{\alpha \beta=1}^{4} w_{i \alpha} \left(\tilde{\tau}_{0} \tilde{\sigma}_{y} \right)_{\alpha,\beta}  w_{i \beta}+\frac{i\gamma}{2} \sum_{i=1}^{N} \sum_{\alpha \beta=1}^{4} w_{i \alpha} \left(\tilde{\tau}_{0} \tilde{\sigma}_{0} + \tilde{\tau}_{x} \tilde{\sigma}_{0} \right)_{\alpha,\beta} w_{i \beta} \label{SubEq:DampingmatrixStep2} .
\end{align}
\end{subequations}
\end{widetext}
Here, $\vect{\tilde{\tau}}$ and $\vect{\tilde{\sigma}}$ denote Pauli matrices representing the dressed degrees of freedom as compared to the original degrees of freedom. Therefore, the Floquet damping matrix $X_F$ can be constructed as
\begin{align}
    U_X(T,0) \equiv e^{i X_F T} = 
    e^{i X_1 \frac{T}{4}} 
    e^{i X_0 \frac{T}{2}} 
    e^{i X_1 \frac{T}{4}} . \label{Eq:DampingFloquetOperator}
\end{align}
Here, $U_X(T,0)$ is a time-evolution operator involving only the Floquet damping matrices. We numerically compute the exponential of the damping matrices $X_0$ [Eq.~\eqref{SubEq:DampingmatrixStep1}] and $X_1$ [Eq.~\eqref{SubEq:DampingmatrixStep2}] to construct the operator $U_X(T,0)$ in Eq.~\eqref{Eq:DampingFloquetOperator}. Afterward, we extract the quasienergies $E_{n}$ and the corresponding eigenstates $\ket{\psi_{n}}$ from the eigenvalue equation: $e^{i X_F T} \ket{\psi_{n}}=\exp(i E_{n} T) \ket{\psi_{n}}$, where the quasienergy $E_{n}$ is now a complex number.

\section{Topological characterization} \label{Subsec:TopoChar}
To extract the topological invariant from $X_F$, we rely on the pseudo-anti-Hermitian symmetry $\Gamma = \tilde{\tau}_x \tilde{\sigma}_x$. This symmetry $\Gamma$ is also called the generalized chiral symmetry for an NH Hamiltonian~\cite{KawabataKPRX2019}. Both the step damping matrices $X_0$ [Eq.~\eqref{SubEq:DampingmatrixStep1}] and $X_1$ [Eq.~\eqref{SubEq:DampingmatrixStep2}] satisfies this symmetry, such that $\Gamma X_i \Gamma = -X_i^\dagger$; with $i=0,1$. The operator $U_X(T,0)$ under the pseudo-anti-Hermitian symmetry transforms as $\Gamma U_X(T,0) \Gamma= U_X^\dagger(T,0)$. We provide this proof in Appendix~\ref{App:FOchiral}. Additionally, the pseudo-anti-Hermitian symmetry divides the full time period into two parts such that: First part $U_c=U_X(T/2,0)$ and second part $\Gamma U_c^\dagger \Gamma=U_X(T,T/2)$~\cite{AsbothPRB2014}. Using these two parts, we can define two operators: $U_X^1=\Gamma U_c^\dagger \Gamma U_c$ and $U_X^2=U_c \Gamma U_c^\dagger \Gamma $. The operators $U_X^{1,2}$ satisfies $\Gamma U^{1,2}_X(T,0) \Gamma= U_X^{1,2\dagger}(T,0)$. Accordingly, we can define two Floquet damping matrices as $X_F^{1,2}=\frac{i}{T} \ln U_X^{1,2}$ with $X_F^{1,2}$ satisfying the pseudo-anti-Hermitian symmetry: $\Gamma X_F^{1,2} \Gamma = -X_F^{1,2\dagger}$. The pseudo-anti-Hermitian symmetry demands the left and right eigenstates ($\ket{\psi_{{\rm L},n}^{1,2}}$ and $\ket{\psi_{{\rm R},n}^{1,2}}$) and the eigenvalues $E_n^{1,2}$ of $X_F^{1,2}$ to satisfy: $\ket{\psi_{{\rm L},-n}^{1,2}}=\Gamma \ket{\psi_{{\rm R},n}^{1,2}}$ and $E_{-n}^{1,2*}=-E_{n}^{1,2}$~\cite{EsakiPRB2011}. Using these properties, we can construct an operator $Q^{1,2}$ as
\begin{align}
    Q^{1,2}=\frac{1}{2} \Bigg(  &\sum_{n>0} \ket{\psi^{1,2}_{{\rm L},n}} \bra{\psi^{1,2}_{{\rm R},n}}  - \sum_{n<0} \ket{\psi^{1,2}_{{\rm L},n}} \bra{\psi^{1,2}_{{\rm R},n}} \non \\
     +&\sum_{n>0} \ket{\psi^{1,2}_{{\rm R},n}}  \bra{\psi^{1,2}_{{\rm L},n}}  - \sum_{n<0} \ket{\psi^{1,2}_{{\rm R},n}}  \bra{\psi^{1,2}_{{\rm L},n}}   \Bigg).
\end{align}
Here, $Q^{1,2}$ is a Hermitian matrix and satisfies $\Gamma Q^{1,2} \Gamma=-Q^{1,2}$, such that $\Gamma$ is an effective chiral symmetry for $Q^{1,2}$. Now, we employ the basis in which $\Gamma$ is diagonal such that $U_\Gamma \Gamma U_\Gamma^{-1}={\rm diag} \left(1,\cdots,1,-1,\cdots,-1 \right)$. In this basis, the matrix $Q^{1,2}$ takes a block anti-diagonal form 
\begin{align}
U_{\Gamma} Q^{1,2} U_{\Gamma}^{-1} = \begin{pmatrix}
0 & q^{1,2} \\
q^{1,2\dagger} & 0
\end{pmatrix}.
\end{align}
Also, the $+1$ and $-1$ eigenvalues effectively provide us with two sublattice degrees of freedom $A$ and $B$, respectively. We can represent $\Gamma$ as $\Gamma=U_A^\Gamma-U_B^\Gamma$; with $U_A^\Gamma=\sum \ket{A} \bra{A}$ and $U_B^\Gamma=\sum \ket{B} \bra{B}$. Now, we perform a singular value decomposition of $q^{1,2}$ as $q^{1,2}=U_A^{1,2}\Sigma^{1,2}U_B^{1,2\dagger}$, with $U_{A,B}^{1,2}$ and $\Sigma^{1,2}$ containing the singular vectors and singular values, respectively. Using $U_{A,B}$ and $U_{A,B}^{1,2}$, we define the winding numbers $\nu_{1,2}$ associated to the Floquet damping matrices $X_F^{1,2}$ as~\cite{LinPRB2021}
\begin{align}
    \nu_{1,2} = \frac{1}{2\pi i} {\rm Tr} \left[\ln \left( \bar{P}_A^{1,2} \bar{P}_B^{1,2\dagger}  \right) \right] ,
\end{align}
where the sublattice dipole moment operators $\bar{P}_A^{1,2}$ and $\bar{P}_B^{1,2}$ are restricted to the sublattice $A$ and $B$, respectively. Explicitly, $\bar{P}_{A,B}^{1,2}$ are defined as $\bar{P}_{A,B}^{1,2}=U_{A,B}^{1,2\dagger} U^{\Gamma\dagger}_{A,B} P U^{\Gamma}_{A,B} U_{A,B}^{1,2}$ with the dipole moment operator $P=\exp(-2\pi i x/N)$. Finally, we combine $\nu_{1,2}$ in such a way that the number of $0$-modes $2 \nu_0$ and $\pi$-modes $2\nu_\pi$ are given as~\cite{AsbothPRB2014}
\begin{align}
    \nu_0 = \frac{\nu_1 + \nu_2}{2}, \quad {\rm and} \quad \nu_\pi = \frac{\nu_1 - \nu_2}{2} .
\end{align}
We use $\nu_{0,\pi}$ to topologically characterize the $0$- and $\pi$-modes that emerges in our system.

\begin{figure*}[t]
    \centering
    \includegraphics[width=0.98\textwidth]{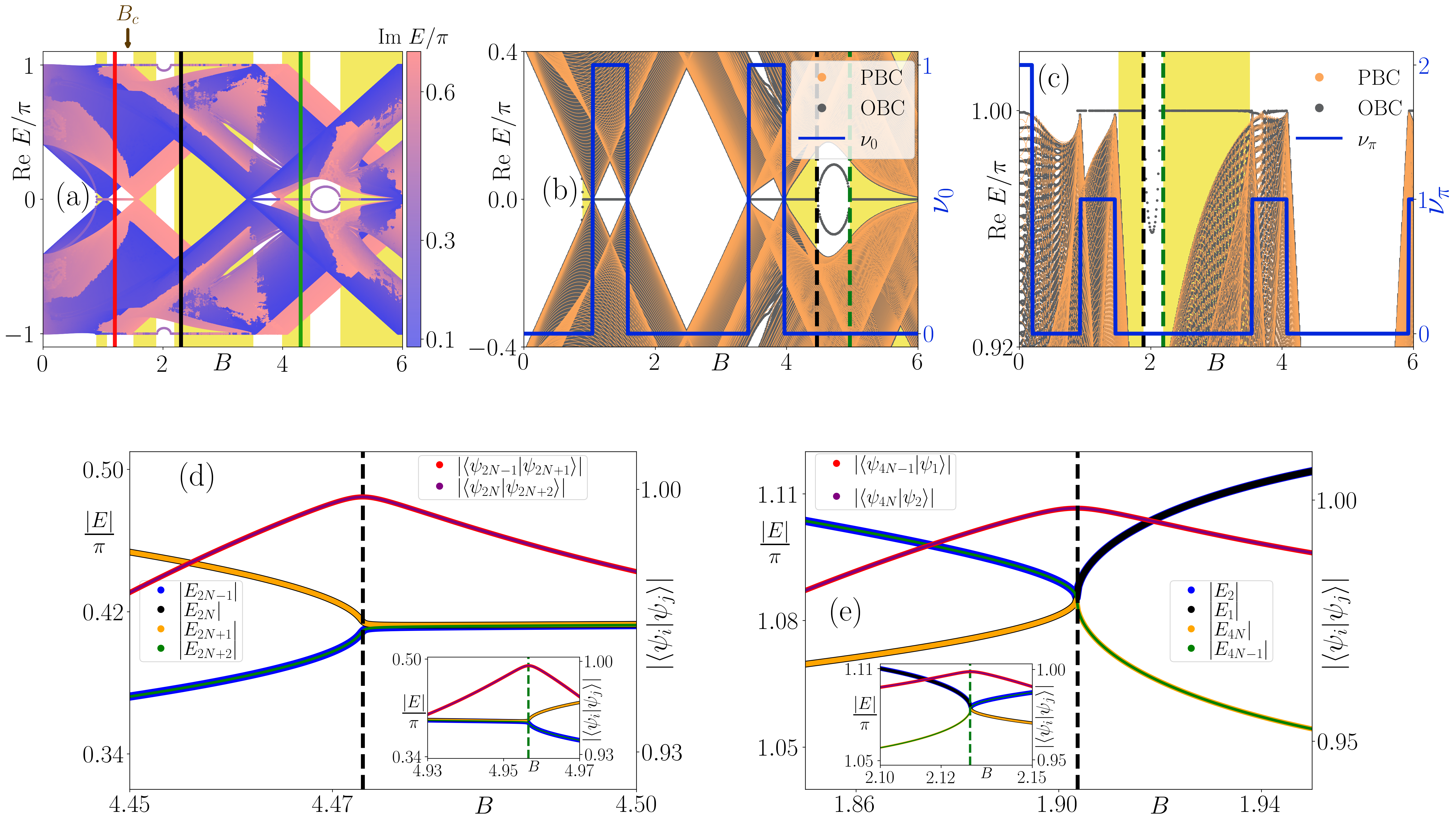}
    \caption{(a) Real part of the quasienergy ${\rm Re}~E$ as a function of the magnetic field $B$ of the Floquet damping matrix, considering OBC. The color represents the imaginary part of the quasienergy ${\rm Im}~E$ and the amplitude is represented by the colorbar. The red, black, and green dashed lines represent parameters used in Fig.~\ref{fig:Eigenvalue_chain_probability_plot}. The yellow shade represents regions of TZMs and TPMs. We use $100$ lattice sites. (b) Quasienergy spectra close to ${\rm Re}~E=0$ for a system obeying PBC (orange) and OBC (gray). The right axis represents the winding number $\nu_0$. We demarcate the equilibrium critical point $B_c=\sqrt{2}$ to highlight the emergence of new topological phases out of a driven dissipative quantum system. We use $400$ lattice sites. (c) We repeat (b) but close to ${\rm Re}~E=\pi$ and the right axis represents the winding number $\nu_\pi$. (d) The absolute quasienergy $\lvert E \rvert$ as a function of $B$ is shown close to the black dashed line in (b). The right axis represents the scalar product $\lvert \langle \psi_i | \psi_j \rangle \rvert$. The inset shows $\lvert E \rvert$ and $\lvert \langle \psi_i | \psi_j \rangle \rvert$ close to the green dashed line in (b). (e) We repeat (d) but for the black and green dashed lines in (c). Here, $2 t_h = \lambda_{R} = \Delta=\gamma=1$, $\mu=0$, $\mu_1=0.3$, and $\omega=2.5$.}
    \label{Fig:PBC-OBC}
\end{figure*}

\section{Results} \label{Sec:Results}
In this section, we discuss the numerical results associated with the driven-dissipative Rashba nanowire. First, we show the emergence of the MZMs and MPMs using the Floquet damping matrix. Note that we always use a finite system obeying open boundary condition~(OBC) while discussing the spectral properties: Eigenvalue spectra and local density of states~(LDOS). To compute the winding numbers, we consider periodic boundary condition~(PBC). We also discuss the TZMs and TPMs and link their emergence to the EPs. Then we study the topological phase diagrams of this driven-dissipative system.

\subsection{Emergence of MZMs and MPMs}
We investigate the quasienergy spectra $E$ of the Floquet damping matrix $X_F$ as a function of $B$ in Fig.~\ref{Fig:PBC-OBC}(a) for a system obeying OBC. The quasienergy $E$ is a complex number in the presence of dissipation, and we show the real part of the quasienergy spectra ${\rm Re}~E$ while the imaginary part ${\rm Im}~E$ is represented by the colormap in Fig.~\ref{Fig:PBC-OBC}(a). The quasienergy spectra reveal the appearance of states near quasienergy ${\rm Re}~E=0,\pm\pi$. 
There are two different kinds of states that appear at these quasienergies. The regions marked by yellow in Fig.~\ref{Fig:PBC-OBC}(a) host modes at quasienergies $0$ and $\pm \pi$, which we
refer to as TZMs and TPMs, respectively. We distinguish the TZMs and TPMs from the MZMs and MPMs in the next subsection. 
These modes at ${\rm Re}~E=0,\pm\pi$ also acquire a finite imaginary part in the dissipative background, implying their state would decay with time.

Figure~\ref{Fig:PBC-OBC}(a) exhibits that the driven system hosts MZMs and MPMs that are connected to the bulk topology of the system. The Majorana modes appearing at quasienergy ${\rm Re}~E=\pm\pi$ are unique to a driven system and have no static analog. Moreover, given the choice of the parameters, the static system is always topologically trivial, such that $B_{c1}=B_{c2}=B_c$ (indicated by an arrow in Fig.~\ref{Fig:PBC-OBC}(a)). Thus, the driven system generates topological MZMs, thereby extending the topological phase boundaries.

To verify that the MZMs and MPMs are directly linked to bulk gap closing and reopening, we also study the quasienergy spectra employing the PBC [orange dots] and compare them with the OBC [gray dots] spectra in Figs.~\ref{Fig:PBC-OBC}(b) and (c), respectively. Note that we have verified that our system is free from the NH skin effect, owing to the particular form of the loss operator in Eq.~\eqref{lossOperator}. Thus, the gap-closing points in PBC spectra coincide with those of OBC, indicating the topological phase boundaries. We focus on the modes that appear at quasienergies $0$ and $\pm \pi$ in Fig.~\ref{Fig:PBC-OBC}(b) and (c), respectively. The MZMs and MPMs appear in the eigenvalue spectra when there is a bulk gap closing in the PBC spectral at ${\rm Re}~E=0$ and $\pm \pi$, respectively. To substantiate our claim that the MZMs and TPMs correspond to topologically non-trivial phases, we calculate the winding number $\nu_{0,\pi}$, shown as the solid blue line in Fig.~\ref{Fig:PBC-OBC}(b,c). The winding number exhibits a clear transition from $\nu_{0}=1$ and $\nu_\pi=1,2$ (topological) to $\nu_{0,\pi}=0$ (trivial) and vice-versa, which occurs through a bulk gap closing. The non-zero values of $\nu_{0,\pi}$ coincide precisely with the regions where MZMs and MPMs appear, thereby confirming that these modes are indeed of topological origin. Also, we numerically check that the MZMs and MPMs are eigenstates of the particle-hole symmetry operator, $\tilde{C}=\tilde{\tau}_0\tilde{\sigma}_0 K$. Thus, the MZMs and MPMs are their own anti-particle, ensuring that they are indeed Majorana modes.

\subsection{Appearance of TZMs and TPMs}
Here, we discuss the TZMs and TPMs that appear at ${\rm Re}~E=0$ and $\pm \pi$, respectively, in the driven-dissipative system. In Fig.~\ref{Fig:PBC-OBC}(a), we mark regions with TZMs and TPMs by a yellow shade. While in Figs.~\ref{Fig:PBC-OBC}(b,c), by analyzing the eigenvalue spectra in OBC and PBC, we observe that the TZMs and TPMs only appear at OBC. Note that the yellow regions in Fig. \ref{Fig:PBC-OBC}(b), and (c) represent the magnetic field windows hosting TZMs and TPMs, respectively. 
Notably, their emergence/ disappearance is not connected to any bulk gap closing-reopening from one side in the PBC spectra. Thus, these modes are denoted as TZMs and TPMs. Furthermore, we also see that the TZMs and TPMs do not carry any non-zero winding numbers $\nu_{0,\pi}$, reassuring that these modes are topologically trivial.

Now, we tie the emergence of the TZMs to the EPs that appear in the spectra~\cite{GhoshDissipation2024}. Although one side of the yellow regions in Figs.~\ref{Fig:PBC-OBC}(a,b), which indicates the TZMs, represents bulk gap closing, the appearance and disappearance of these modes are not directly linked to the bulk gap opening and reopening. Here, we discuss how the EPs play a crucial role in the emergence and destruction of the TZMs. First, we focus on the regions that appear close to $B=4$ in Fig.~\ref{Fig:PBC-OBC}(b). The left side of this yellow region is a bulk gap closing point. However, the TZMs in this region can be destroyed without closing any bulk gap as happens at the point indicated by the black dashed line in Fig.~\ref{Fig:PBC-OBC}(b) [$B \simeq 4.473$]. For this particular value of the magnetic field $B$, we observe the formation of two second-order exceptional points (EPs), as shown in Fig.~\ref{Fig:PBC-OBC}(d). Specifically, the $(2N-1)$-th (blue) state coalesces with the $(2N+1)$-th (orange) state, while the $2N$-th (black) state coalesces with the $(2N+2)$-th (green) state, leading to two distinct second-order EPs. At these points, the corresponding eigenvalues become degenerate in their real, imaginary, and absolute parts, and the associated eigenstates also coalesce. In Fig.~\ref{Fig:PBC-OBC}(d), we plot only the absolute eigenvalues $|E|$ as a function of $B$, but both the real and imaginary components exhibit the same coalescence behavior. The eigenvalues are ordered according to their real parts. To confirm the coalescence of the eigenstates, we compute the scalar product between the relevant pairs, shown on the right axis of Fig.~\ref{Fig:PBC-OBC}(d). We find that $|\langle \psi_{2N-1} | \psi_{2N+1} \rangle|$ (red) and $|\langle \psi_{2N} | \psi_{2N+2} \rangle|$ (purple) both approach unity, indicating that the corresponding eigenstates become parallel at the EPs. The black dashed line in Fig.~\ref{Fig:PBC-OBC}(b) therefore marks the locations of these second-order EPs, whose emergence leads to the destruction of the trivial zero modes (TZMs).
On the other hand, the TZMs can also be created by forming EPs, which happens at the green dashed line in Fig.~\ref{Fig:PBC-OBC}(b). At this point, we see the formation of two second-order EPs, which are shown in the inset of Fig.~\ref{Fig:PBC-OBC}(d). In the literature, the EPs have been shown to have a topological nature and can be connected to energy-based topological indices~\cite{TonyLeePRL2016,Hodaei2017,Ghatak2019,AshidaYotoAP2020,DennerNatComm2021,BergholtzRMP2021,SayyadPRR2022,OkumaAnnualRev2023,BanerjeeJPCM2023}. Here, we demonstrate that the EPs can also create and destroy trivial edge modes.

Furthermore, we also relate the emergence and destruction of TPMs to the EPs. Note that the yellow shaded region is bounded by EP and bulk gap closing. In this regard, we focus on the black and green dashed line in Fig.~\ref{Fig:PBC-OBC}(c), which indicates the destruction and emergence of the TPMs. Focusing on these lines, we show the formation of two second-order EPs as a function of $B$ in Fig.~\ref{Fig:PBC-OBC}(e). The $(4N-1)$-th (green) state coalesces with the $1$-st (black) state, and the $4N$-th (orange) state coalesces with the $2$-nd (blue) state, forming two second-order EPs. Note that the $1$-st and $2$-nd eigenstates appear at quasienergy $-\pi$, but in a Floquet system, the quasienergy $-\pi$ is equivalent to quasienergy $+\pi$. The scalar product of the eigenstates (shown on the right axis of Fig.~\ref{Fig:PBC-OBC}(e)) $\lvert \langle \psi_{4N-1}| \psi_{1} \rangle\rvert$ (red) and $\lvert \langle \psi_{4N}| \psi_{2} \rangle\rvert$ (purple) also becomes unity, indicating that these states become pairwise parallel to each other. Thus, the black dashed line in Fig.~\ref{Fig:PBC-OBC}(c) indicates EPs. Similarly, at the green dashed line in Fig.~\ref{Fig:PBC-OBC}(c), we also obtain two second-order EPs, and their formation is shown in the inset of Fig.~\ref{Fig:PBC-OBC}(e). In summary, the EPs play a key role in the formation of the TZMs and TPMs in a driven system, demonstrating its significance in shaping the system's dynamics and spectral properties. Our observation is not limited by any specific choice of parameters and is a property of the driven Rashba nanowire in the presence of the dissipation of the form given in Eq.~\eqref{lossOperator}.

\begin{figure}
    \centering
    \includegraphics[width=0.48\textwidth]{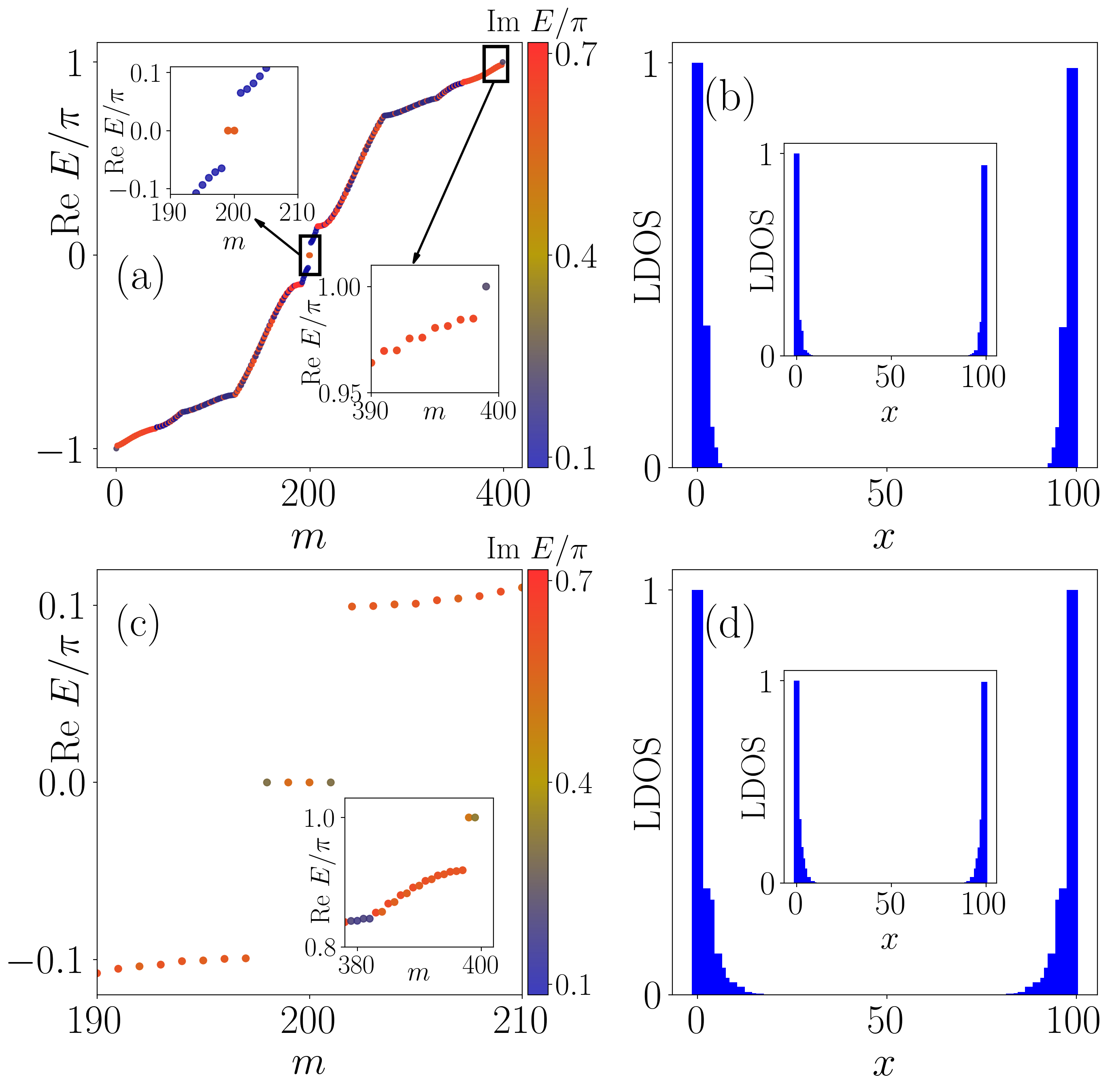}
    \caption{(a) Real part of the quasienergy spectrum ${\rm Re}~E$ as a function of $m$ for $B=1.2$ (red line in Fig.~\ref{Fig:PBC-OBC}(a)). The insets represent zoomed-in spectra close to ${\rm Re}~E=0$ and $\pi$. The color bar indicates the imaginary part of the quasienergy. (b) LDOS associated with the MZMs, while the inset shows the LDOS for the MPMs. (c) Real part of the quasienergy spectrum close to ${\rm Re}~E=0.0$ showing four TZMs for $B=4.3$ (green line in Fig.~\ref{Fig:PBC-OBC}(a)). The inset shows real part of the quasienergy spectrum close to ${\rm Re}~E=\pi$, showing two TPMs (the other two appear at ${\rm Re}~E=-\pi$) for $B=2.3$ (black line in Fig.~\ref{Fig:PBC-OBC}(a)). (d) LDOS associated with the TZMs. The inset shows the LDOS for the TPMs. We consider $N=100$ lattice sites and the rest of the parameters remain the same as in Fig.~\ref{Fig:PBC-OBC}.}
    \label{fig:Eigenvalue_chain_probability_plot} 
\end{figure}

\subsection{Local density of states}

Having discussed the emergence of different kinds of $0$ and $\pi$-modes in the system, we now demonstrate their localization properties in Fig.~\ref{fig:Eigenvalue_chain_probability_plot}. To this end, we focus on some specific parameters as indicated by the red, black, and green lines in Fig.~\ref{Fig:PBC-OBC}(a). For the red line in Fig.~\ref{Fig:PBC-OBC}(a), we show the real part of eigenvalue spectrum ${\rm Re}~E$ of the Floquet damping matrix $X_F$ obeying OBC as a function of the eigenvalue index $m$ in Fig.~\ref{fig:Eigenvalue_chain_probability_plot}(a). The color bar represents the imaginary part of the eigenvalue ${\rm Im}~E$. In the insets, we show zoomed-in spectra close to ${\rm Re}~E=0$ and ${\rm Re}~E=\pi$, which exhibit the presence of two MZMs and MPMs, respectively. Note that for this parameter, the system also exhibits a non-zero winding number $\nu_{0,\pi}=1$. On the other hand, both the MZMs and MPMs have a non-zero imaginary part, as seen from the color of the eigenvalues. Thus, the MZMs and MPMs have a finite lifetime. We compute the local density at ${\rm Re}~E=0$ and $\pm \pi$ and show them in Fig.~\ref{fig:Eigenvalue_chain_probability_plot}(b) and its inset, respectively, which show the edge localization of the Majorana modes.

To focus on the TZMs and TPMs, we now consider the green and black lines in Fig.~\ref{Fig:PBC-OBC}(a). For the green~(black) line, the system exhibits four modes at quasienergy ${\rm Re}~E=0$~(two modes each at quasienergy ${\rm Re}~E=+\pi$ and $-\pi$), which is demonstrated in Fig.~\ref{fig:Eigenvalue_chain_probability_plot}(c) (inset of Fig.~\ref{fig:Eigenvalue_chain_probability_plot}(c)). These modes, however, do not carry any winding number \ie $\nu_{0,\pi}=0$. Hence, they are denoted as TZMs and TPMs. Nevertheless, the TZMs and TPMs are also located at the edges of the system, as shown in Fig.~\ref{fig:Eigenvalue_chain_probability_plot}(d) and its inset, respectively. Note that our system is free from the NH skin-effect; thus, the boundary modes that appear in our system are either MZMs, MPMs, TZMs, or TPMs.

In summary, we demonstrate that the drive-dissipative system hosts topologically non-trivial MZMs and MPMs, which are topologically characterized by non-zero winding numbers. The interplay of dissipation and periodic drive also gives rise to trivial boundary modes both at quasienergy $0$ and $\pi$, which are not connected to any bulk topology; rather, their creation and destruction are linked to the EPs. The MPMs and TPMs only appear in a driven-dissipative system and have no static analog~\cite{GhoshDissipation2024}. In Appendix~\ref{App:Disorder}, we discuss the effect of onsite disorder on the boundary modes. Our finding suggests that the MZMs and MPMs, as well as the TZMs and TPMs, are robust against disorder.

\begin{figure}
    \centering
    \includegraphics[width=0.48\textwidth]{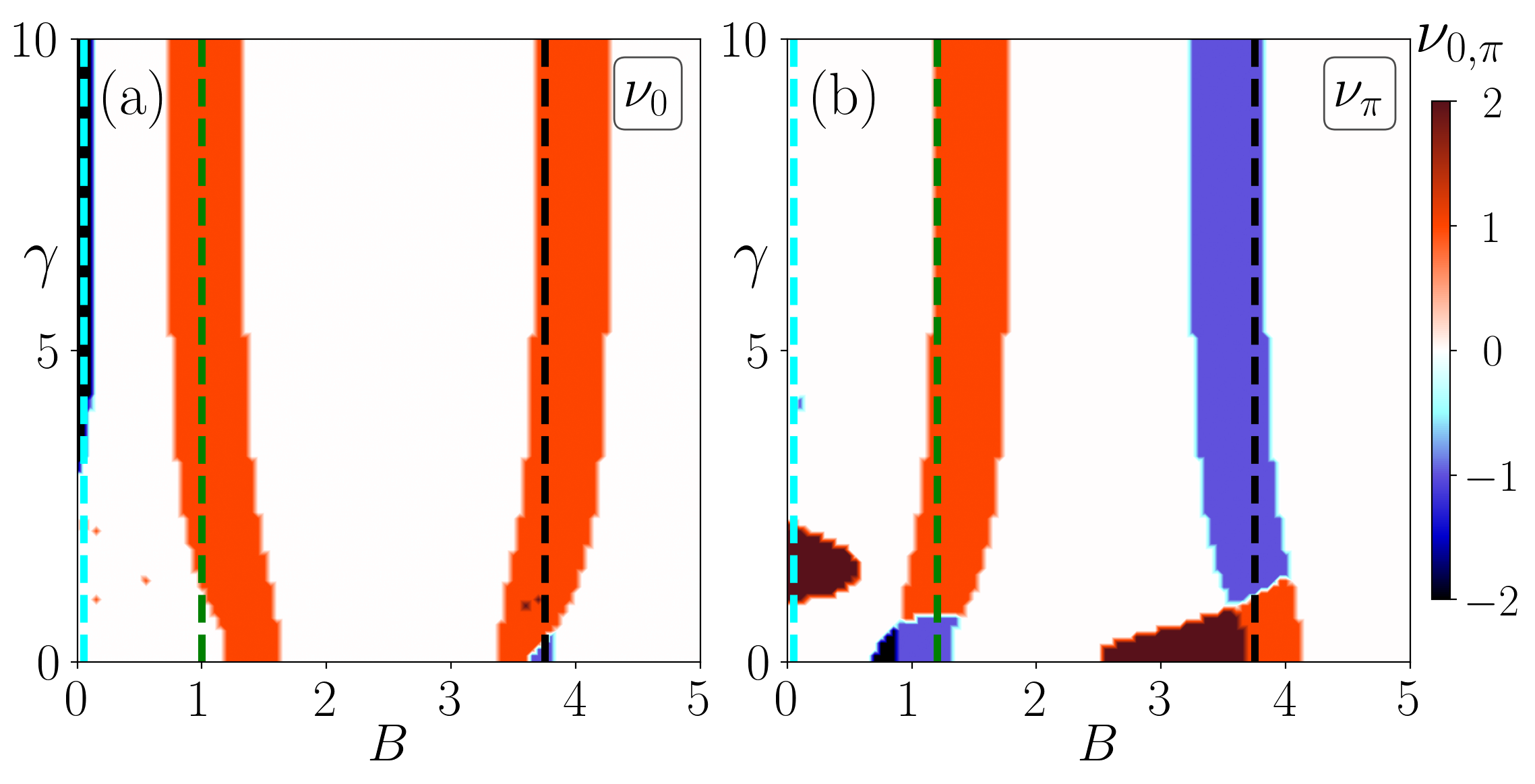}
    \caption{The Phase diagram in terms of (a) $\nu_0$ and (b) $\nu_\pi$ in the $B \mhyphen \gamma$ plane. The $\gamma=0$ line represents the Hermitian driven system. Here, the colorbar represents the values of $\nu_{0,\pi}$. We consider $200$ lattice sites. Rest of the parameters remain the same as in Fig.~\ref{Fig:PBC-OBC}.}
    \label{Fig:Phasediagram}
\end{figure}

\subsection{Phase diagram and higher winding numbers}
Having discussed the spectral properties of the system, we now investigate the phase diagram of this driven-dissipative system in Fig.~\ref{Fig:Phasediagram}. In particular, we show the winding number $\nu_0$ and $\nu_\pi$ in the $B \mhyphen \gamma$ plane in Figs.~\ref{Fig:Phasediagram}(a) and (b), respectively. Focusing on $\nu_0$ in Fig.~\ref{Fig:Phasediagram}(a), we observe phases with winding numbers $\lvert \nu_{0} \rvert > 1$, which indicates the appearance of multiple (more than two) edge-localized MZMs. Generation of multiple topological modes is one of the salient features of the driven system~\cite{MondalPRB2023}. In Appendix~\ref{App:Higer_winding}, we explicitly demonstrate the eigenvalue spectra and the LDOS associated with the higher winding number phases.
Furthermore, if we compare the $\nu_0$ between $\gamma=0$ (Hermitian case) and $\gamma>0$ (non-Hermitian case), we observe a phase transition with respect to $\gamma$. To further elucidate this, we draw a few vertical lines in Fig.~\ref{Fig:Phasediagram}(a): cyan ($B=0.05$), green ($B=1.0$), and black ($B=3.75$). The cyan and green lines indicate a phase transition as a function of $\gamma$ for a magnetic field $B<B_{c}(=\sqrt{2})$. Thus, in a driven-dissipative system, we can obtain the MZMs for a lesser strength of applied magnetic field compared to the static system, which can be advantageous for an experimental setup. Furthermore, the black dashed line reveals a change in winding number $\nu_0=-1\rightarrow1$ as a function of the dissipation strength $\gamma$. Thus, the phase diagram Fig.~\ref{Fig:Phasediagram}(a) indicates that the dissipation can substantially modify the topological phases of the driven system and induce topological states which are only realized in the presence of dissipation~\cite{GhoshDissipation2024}. This is one of the key observations of this work.

We also obtain a rich phase diagram for the MPMs, see Fig.~\ref{Fig:Phasediagram}(b), where we show $\nu_\pi$ in the $B \mhyphen \gamma$ plane. As mentioned before, the generation of the MPMs is unique to a driven system without any static analog. Figure~\ref{Fig:Phasediagram}(b) exhibits phases with a higher winding number $\lvert \nu_\pi \rvert >  1$, which indicates generation of multiple MPMs~\cite{MondalPRB2023}. Moreover, Fig.~\ref{Fig:Phasediagram}(b) suggests the topological phase transition with respect to $\gamma$, and also, there are some phases that do not exist in the closed driven system, i.e., the Hermitian case. To investigate this further, we draw a few vertical lines: cyan ($B=0.5$), green ($B=1.2$), and black ($B=3.75$). These lines demonstrate the change in winding number $\nu_\pi$ as a function of $\gamma$ and show how dissipation changes the topology of the system compared to a Hermitian-driven system. Together in Fig.~\ref{Fig:Phasediagram}, we uncover the important role of dissipation in shaping the topological phases in a driven system.

\section{Possible experimental framework} \label{Sec:Exp}
In this section, we discuss the possible experimental feasibility of our work. The heterostructure setup can be realized using semiconducting ${\rm InSb/InAs}$ nanowire, which has a strong SOC and ${\rm Nb/Al}$ as the superconducting material to induce the $s$-wave gap in the nanowire via the proximity effect~\cite{das2012zero,Mourik2012Science,NichelePRL2017,JunSciAdv2017,Zhang2017NatCommun,Gul2018,Grivnin2019,ChenPRL2019}.
The SOC can be further enhanced by employing an electrostatic gate~\cite{ReutherPRX2013} and also using interfaces~\cite{Wangnc2015}. The dissipation would arise from the coupling to a bath, which can be a metallic lead close to the nanowire.
In our work, we have considered a loss that is onsite in nature. However, in a real experimental setup, the loss can have a complicated form, such as hopping losses. Thus, considering different types of jump operators might also be very intriguing.

The step-drive protocol can be implemented using a time-dependent gate voltage by superposing several harmonics to obtain a sharp step-like jump~\cite{khosravi2009bound,dubois2013minimal,Gabelli13,Misiorny18}. The step Hamiltonian $H_1$, which represents an atomic insulator and occurs in the first and last step in Eq.~\eqref{Eq:Drive}, can be realized by tuning the gate voltage in such a way that the system exhibits almost flat and gapped flat bands. In contrast, the Hamiltonian $H_0$ in the time interval $T/4 < t < 3T/4$ can be obtained by adjusting the gate voltage such that the band dispersion is relevant, thereby achieving the nanowire limit. Experimentally implementing a very sharp step-drive is challenging, and it is most likely to have a short switching time $t_s$. In our system, the other relevant time-scales are $\xi_M/v$ ($\xi_M$ is the MZM/MPM localization length and $v$ is the Fermi velocity of the quasi particles) and $1/\Delta E_M$ ($\Delta E_M$ denotes the gap separating the MZMs/MPMs from other states). The first condition prevents the MZMs/MPMs from hybridizing with the bulk states, and they remain localized. At the same time, the second condition ensures a non-adiabatic evolution, allowing us to engineer the MZMs/MPMs. This is important since we begin with an atomic insulator in the first step, $H_1$, and an adiabatic evolution would restrict the formation of the non-trivial MZMs/MPMs. Thus, the switching timescale $t_s< {\rm min}(\xi_M/v,1/\Delta E_M)$ to successfully generate the MZMs and MPMs in our setup.

To provide some experimentally relevant parameters for our model, we consider the Rashba SOC strength of ${\rm InSb}$ nanowire, which is around $50~\mu{\rm eV}$~\cite{Mourik2012Science}. Thus, the model parameters for our system can take the values as follows $\lambda_{R} \simeq 50~\mu{\rm eV}$, $t_h \simeq 25~\mu{\rm eV}$, $\Delta \simeq 50~\mu{\rm eV}$, and $B\simeq 20~{\rm mT}$.

\section{Summary and Discussion} \label{Sec:Summary}
In this work, we consider a Rashba nanowire in close proximity to an $s$-wave superconductor. We use a periodic step-drive and assume an onsite loss in the system. The dynamics of the system is governed by a periodic Lindbladian. We employ the third-quantization method to obtain the Floquet Lindbladian and thereby the Floquet damping matrix that provides the spectral and topological properties of the system. Our system is free from skin-effect for this choice of jump operator. We investigate the spectra of the Floquet damping matrix and show the existence of the MZMs and MPMs. The MZMs and MPMs, however, get a finite imaginary part in the presence of dissipation. Additionally, the driven-dissipative system also hosts TZMs and TPMs, whose origin is not connected to the bulk topology of the system; rather, they are induced by the EPs. We compute the topological invariant based on pseudo-anti-Hermitian symmetry to show that the MZMs and MPMs are indeed topological. We also study the localization properties of the MZMs, MPMs, TZMs, and TPMs, and establish their edge localization. Furthermore, we utilize the topological invariant to obtain a phase diagram for this driven-dissipative system. The phase diagram suggests that the dissipation plays an important role in shaping the topological nature of the system and even induces topological phases in the system that are absent in the closed driven system.

In this work, we only consider a 1D Rashba nanowire. However, the MZMs and MPMs can also appear in higher dimensions, such as the corner modes of a two-dimensional second-order topological superconductor~(SOTSC) and a three-dimensional third-order topological superconductor~(TOTSC)~\cite{GhoshDynamical2022}. Thus, one of the interesting outlooks can be to investigate the effect of dissipation in a driven SOTSC and TOTSC and see if the dissipation can induce the MZMs and MPMs in these systems. Also, it would be fascinating to explore if one can also obtain the TZMs and TPMs as the corner modes in these systems.

\subsection*{Data Availability}
The data that support the findings of this article are openly available at Ref.~\cite{GogoiData}.

\acknowledgments
K.G. and T.N. acknowledge the NFSG ``NFSG/HYD/2023/H0911'' from BITS Pilani.

\appendix
\newcounter{defcounter}
\setcounter{defcounter}{0}

\vspace{-1cm}

\section{Procedure to obtain $\mathcal{L}_+$} \label{App:LiouvillianApp}
Here we provide the detailed steps to obtain $\mathcal{L}_+$ [Eq.~(\ref{Liouvillianplus})] using the adjoint fermion operators $\hat{\phi}$'s. The adjoint fermion operators follow the following relations with the basis states $\ket{P_{\underline{\alpha}}}$
\begin{align}
    \ket{w_a P_{\underline{\alpha}}}=& \left( \hat{\phi}_a^\dagger+ \hat{\phi}_a\right) \ket{P_{\underline{\alpha}}}\ , \non \\
    \ket{ P_{\underline{\alpha}}w_a}=& \mathcal{P}_{\rm F} \left( \hat{\phi}_a^\dagger- \hat{\phi}_a\right) \ket{P_{\underline{\alpha}}},
    \label{idnty1}
\end{align}
where $\mathcal{P}_{\rm F}=(-1)^{\sum_a \alpha_a}$ and it represents the fermion parity operator. Furthermore, using Eqs.~(\ref{idnty1}) and the anti-commutation relations of the adjoint fermion operators, we can also derive the following identities
\begin{subequations}
    \begin{align}
    \ket{w_a w_b P_{\underline{\alpha}}}=& \left( \hat{\phi}_a^\dagger\hat{\phi}_b^\dagger +\hat{\phi}_a^\dagger\hat{\phi}_b +\hat{\phi}_a\hat{\phi}_b^\dagger +\hat{\phi}_a\hat{\phi}_b\right) \ket{P_{\underline{\alpha}}} \ , 
    \label{idt1}
    \\
    \ket{P_{\underline{\alpha}} w_a w_b }=& \left( \hat{\phi}_a^\dagger\hat{\phi}_b^\dagger -\hat{\phi}_a^\dagger\hat{\phi}_b -\hat{\phi}_a\hat{\phi}_b^\dagger +\hat{\phi}_a\hat{\phi}_b + 2 \delta_{a,b} \right) \ket{P_{\underline{\alpha}}} \ ,
    \label{idt2}
    \\
    \ket{w_a P_{\underline{\alpha}}  w_b }=& \mathcal{P}_{\rm F} \left( \hat{\phi}_a^\dagger\hat{\phi}_b^\dagger -\hat{\phi}_a^\dagger\hat{\phi}_b +\hat{\phi}_a\hat{\phi}_b^\dagger -\hat{\phi}_a\hat{\phi}_b\right) \ket{P_{\underline{\alpha}}} \ .
    \label{idt3} 
\end{align}
\end{subequations}

Now, we represent $\rho(t)$ using the basis element $\ket{P_{\underline{\alpha}}}$. Using Eqs.~(\ref{idt1}),~(\ref{idt2}), and $H_{M}^{ba}=-H_{M}^{ab}$, the unitary part, i.e., the first term of the Liouvillian $\mathcal{L}$ in Eq.~\eqref{Lindblad1} can be written as
\begin{align}
    i \left[ H, \rho (t) \right] &= -i\sum_{a,b} H_{a,b} \left( \ket{w_a w_b P_{\underline{\alpha}}} -  \ket{P_{\underline{\alpha}} w_a w_b }\right) \non \\
&= -4i\sum_{a,b}  \hat{\phi}_a^\dagger H_{a,b} \hat{\phi}_b \non \\
&=-4i \underline{\phi}^\dagger \cdot H_{ M} \underline{\phi} . \label{Eq:Unitarypart}
\end{align}
The non-unitary part, i.e., the second term in Eq.~\eqref{Lindblad1} can be represented as
\begin{widetext}
\begin{align}
   & - \frac{1}{2} \sum_{m} \left( \left\{ L^\dagger_m L_m , \rho(t) \right\}-2 L^\dagger_m \rho(t) L_m \right) 
   = - \frac{1}{2} \sum_{m,a,b} l_{m,a}l_{m,b}^{*} \left( \ket{w_a w_b P_{\underline{\alpha}}} +\ket{ P_{\underline{\alpha}}w_a w_b} - 2 \ket{w_a  P_{\underline{\alpha}}w_b} \right) \non \\
   =&\sum_{m,a,b} \frac{1+\mathcal{P}_{\rm F}}{2} \left[ \hat{\phi}_a^\dagger \left(l_{m,a}l_{m,b}^*-l_{m,b}l_{m,b}^* \right) \hat{\phi}_b^\dagger - \hat{\phi}_a^\dagger \left(l_{m,a}l_{m,b}^*+l_{m,b}l_{m,b}^* \right) \hat{\phi}_b \right] \non \\
   &+\sum_{m,a,b} \frac{1-\mathcal{P}_{\rm F}}{2} \left[ \hat{\phi}_a \left(l_{m,a}l_{m,b}^*-l_{m,b}l_{m,b}^* \right) \hat{\phi}_b - \hat{\phi}_a \left(l_{m,a}l_{m,b}^*+l_{m,b}l_{m,b}^* \right) \hat{\phi}_b^\dagger \right] \non \\
   & \qquad \qquad \left({\rm using~Eqs.~(\ref{idt1}),~(\ref{idt2}),~and~(\ref{idt3})}\right) \non \\
   =&\sum_{a,b} \frac{1+\mathcal{P}_{\rm F}}{2} \left[ \hat{\phi}_a^\dagger \left(M_{a,b}-M_{b,a} \right) \hat{\phi}_b^\dagger - \hat{\phi}_a^\dagger \left(M_{a,b}+M_{b,a} \right) \hat{\phi}_b \right] +\sum_{a,b} \frac{1-\mathcal{P}_{\rm F}}{2} \left[ \hat{\phi}_a \left(M_{a,b}-M_{b,a} \right) \hat{\phi}_b - \hat{\phi}_a \left(M_{a,b}+M_{b,a} \right) \hat{\phi}_b^\dagger \right] \non \\
   &\qquad \left({\rm here~}M_{a,b}=\sum_m l_{m,a}l_{m,b}^*  \right) \non \\
   =&\frac{1+\mathcal{P}_{\rm F}}{2} \left[ \underline{\phi}^\dagger \cdot \left(M-M^T \right) \underline{\phi}^\dagger - \underline{\phi}^\dagger \cdot \left(M+M^T \right) \underline{\phi} \right] + \frac{1-\mathcal{P}_{\rm F}}{2} \left[ \underline{\phi} \cdot \left(M-M^T \right) \underline{\phi} - \underline{\phi} \cdot \left(M+M^T \right) \underline{\phi}^\dagger \right] \ . \label{Eq:Non-unitarypart}
\end{align}
Therefore, using Eqs.~\eqref{Eq:Unitarypart} and \eqref{Eq:Non-unitarypart}, we can write the full Liouvillian operator as 
\begin{align}
    \mathcal{L}=&-4i \underline{\phi}^\dagger \cdot H_{M} \underline{\phi}+\frac{1+\mathcal{P}_{\rm F}}{2} \left[ \underline{\phi}^\dagger \cdot \left(M-M^T \right) \underline{\phi}^\dagger - \underline{\phi}^\dagger \cdot \left(M+M^T \right) \underline{\phi} \right] + \frac{1-\mathcal{P}_{\rm F}}{2} \left[ \underline{\phi} \cdot \left(M-M^T \right) \underline{\phi} - \underline{\phi} \cdot \left(M+M^T \right) \underline{\phi}^\dagger \right] \ .
   \label{fullLiouvillian}
\end{align}
The physical situation represents a system with an even number of Majorana fermionic operators, such that the total fermionic parity is $\mathcal{P}_{\rm F}=+1$. Thus, in Eq.~(\ref{fullLiouvillian}), only the first term is non-zero and we arrive at $\mathcal{L}_+$ given in Eq.~(\ref{Liouvillianplus}). 
\end{widetext}

\section{$U_X(T,0)$ under chiral symmetry} \label{App:FOchiral}
Here, we discuss how the operator $U_X(t,0)$ transforms under pseudo-anti-Hermitian symmetry $\Gamma$. The damping matix $X(t)$ transforms under pseudo-anti-Hermitian symmetry
as $\Gamma X(t) \Gamma=-X^\dagger(T-t)$. Employing this property, we can obtain the effect on $U_X(t,0)$ as
\begin{align}
    \Gamma  U_X & (t,0) \Gamma \non \\
    =& \Gamma  \left[  {\rm TO} \exp \left( -i \int_0^t X(t') dt' \right) \right] \Gamma  \non \\
    =& \sum_m \frac{(-i)^m}{m!} {\rm TO} \! \! \int_0^t \! dt_1'  \cdots  \int_0^t dt_m' \Gamma X(t_1')\Gamma \ \cdots \ \Gamma X(t_m')\Gamma \non \\
    =& \sum_m \! \frac{(i)^m}{m!} {\rm TO} \! \!\int_0^t \! dt_1'   \cdots  \int_0^t \! dt_m' X^\dagger(T-t_1')  \cdots  X^\dagger(T-t_m') \non \\
    =& \sum_m \frac{(-i)^m}{m!} {\rm TO} \int_T^{T-t} dt_1'   \cdots  \int_T^{T-t} dt_m' X^\dagger(t_1') \cdots X^\dagger(t_m') \non \\
    =& {\rm TO} \exp \left( -i \int_T^{T-t} X^\dagger(t') dt' \right) .
\end{align}
Thus, at the full time-period $T$, the operator $U_X(T,0)$ transforms $ \Gamma U_X(T,0) \Gamma = {\rm TO} \exp \left( -i \int_T^{0} X^\dagger(t') dt' \right)\equiv U^\dagger_X(T,0) $.

\section{Effect of disorder on the Majorana and trivial modes} \label{App:Disorder}

\begin{figure}
    \centering
    \includegraphics[width=0.48\textwidth]{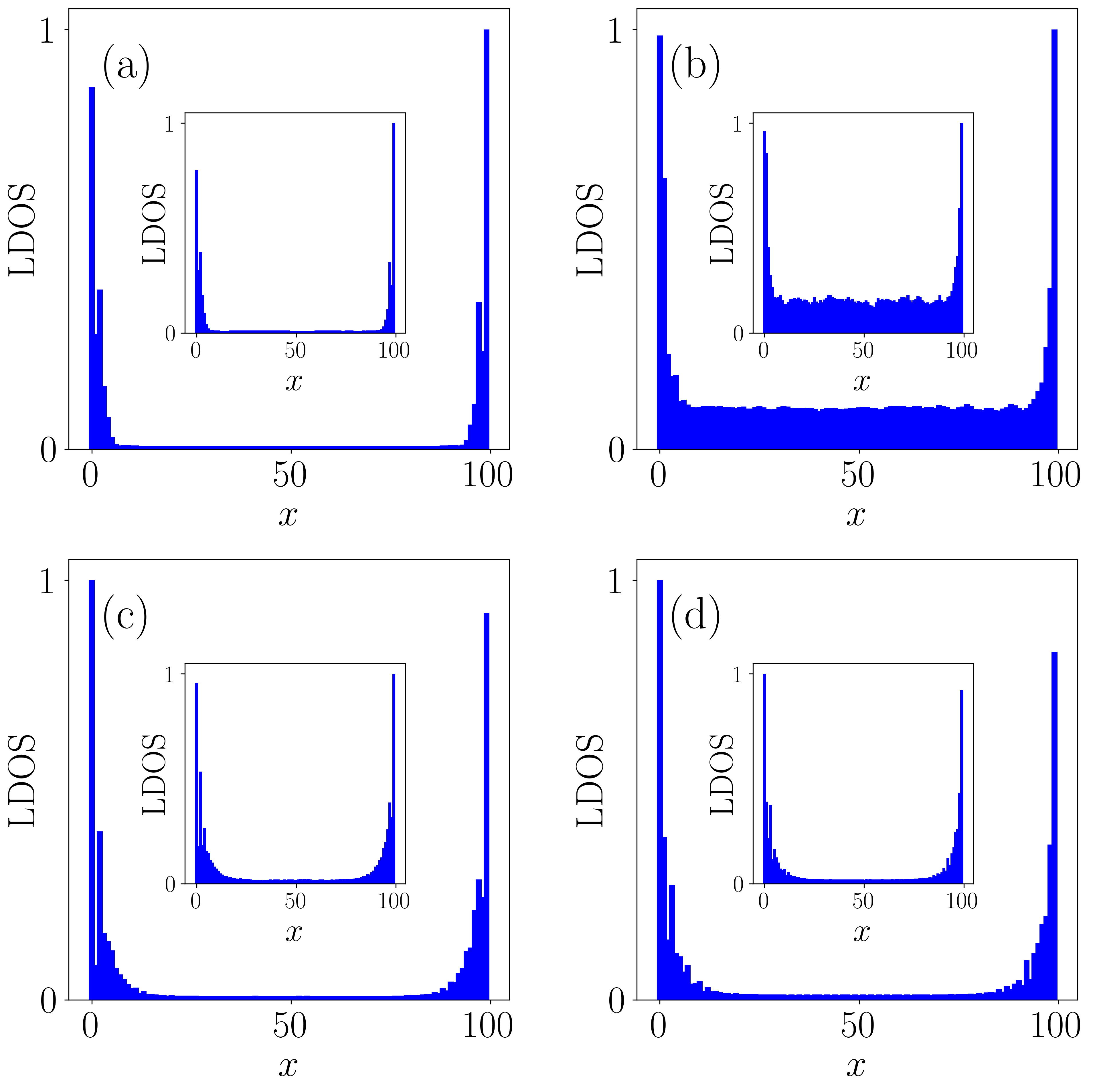}
    \caption{Disorder-averaged LDOS for (a) MZMs, (b) MPMs, (c) TZMs, and (d) TPMs for $W=0.25$. The inset shows LDOS associated with $W=0.5$. We consider $100$ independent disorder realizations for computing disorder averaging. The parameters remain the same as Fig.~\ref{fig:Eigenvalue_chain_probability_plot}. }
   \label{fig:figure_disorder}
\end{figure}

We investigate the robustness of the Majorana (MZMs and MPMs) and the trivial (TZMs and TPMs) modes against onsite random Anderson-type disorder. To this end, we introduce the following term in the step Hamiltonians of Eqs.~\eqref{SubEq:DampingmatrixStep1} and \eqref{SubEq:DampingmatrixStep2}: $H_{\rm dis}=\sum_{i}^N \sum_{\alpha\beta=1}^4 V_i w_{i \alpha} \left( \tilde{\tau}_0 \tilde{\sigma_y} \right)_{\alpha,\beta} w_{i \beta} $. Here, $V_i$ is uniformly distributed in the range $[-W/2,\, W/2]$ with $W$ denoting the disorder strength. We consider $100$ random disorder configurations and take an average over the number of disorder configurations to obtain the disorder-averaged LDOS, as shown in Fig.~\ref{fig:figure_disorder}.

Figures~\ref{fig:figure_disorder}(a) and (b) exhibit the disorder-averaged LDOS associated with MZMs and MPMs, respectively, for a disorder strength $W=0.25$, while the insets show the results for $W=0.5$. The MZMs and MPMs survive and are localized at the edges of the system for both disorder strengths, implying that they are robust against disorder. In Figs.~\ref{fig:figure_disorder}(c) and (d), we illustrate the disorder-averaged LDOS for the TZMs and TPMs for $W=0.25$, and the insets show the same but for $W=0.5$.
Surprisingly, the TZMs and TPMs also remain localized at the edges of the system, even though they are not protected by any bulk topology of the system~\cite{GhoshDissipation2024}.

\section{Higher winding number phase: Multiple MZMs and MPMs} \label{App:Higer_winding}

\begin{figure}
    \centering
    \includegraphics[width=0.48\textwidth]{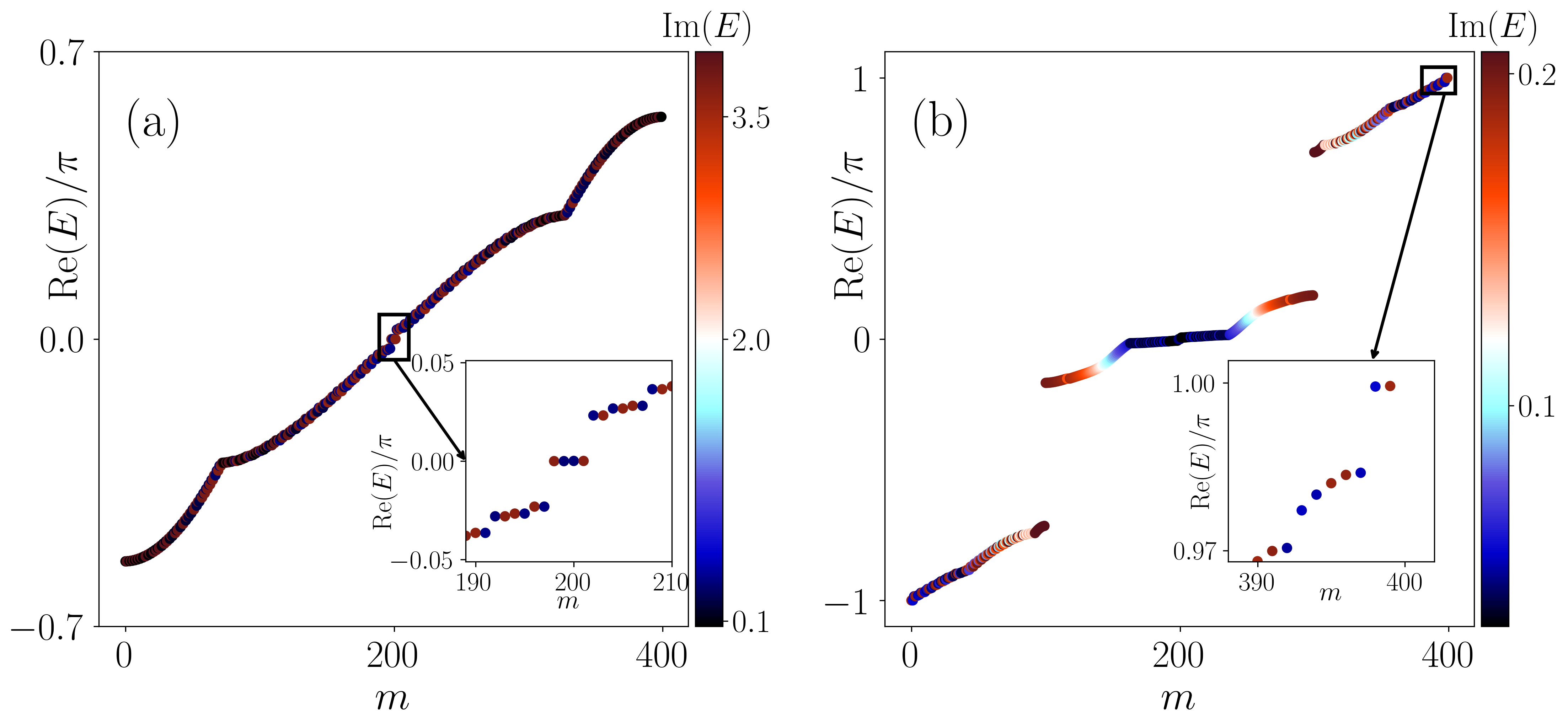}
    \caption{Real part of the quasienergy spectrum as a function of the state index showing (a)  $\nu_0=2$ phase ($B=0$ and $\gamma=5$), (b) $\nu_\pi=2$ phase ($B=3.4$ and $\gamma=0.3$). The insets show zoomed-in spectra around ${\rm Re}~E = 0$ in (a) and around ${\rm Re}~E = \pi$ in (b). The colorbar represents the imaginary part of the eigenvalue. The rest of the parameters take the same value as Fig.~\ref{Fig:PBC-OBC}.}
    \label{Fig:higher_winding}
\end{figure}

In Fig.~\ref{Fig:Phasediagram}, we show phases with higher winding numbers $\nu_{0,\pi} > 1$, which suggests the existence of multiple MZMs and MPMs. Here, we explicitly demonstrate the spectral signatures of the MZMs and MPMs in terms of the quasienergy spectrum. In Fig.~\ref{Fig:higher_winding}(a), we show the quasienrgy spectrum as a function of the state index corresponding to the phase with $\nu_0=2$, and a zoomed-in quasienergy spectrum is shown in the inset. Figure~\ref{Fig:higher_winding}(a) and its inset exhibit that there are four MZMs at quasienrgy $E=0$. We check that these four MZMs are localized at the edges of the system (not shown here, but resembling Fig.~\ref{fig:Eigenvalue_chain_probability_plot}(a)).

In Fig.~\ref{Fig:higher_winding}(b), we show the quasienergy spectrum for a case with $\nu_\pi=2$. The quasienergy spectrum shows the presence of four MPMs (two at each of the quasienergies $\pm \pi$). Again, we check that the MPMs are localized at the edges of the system. The generation of multiple Majorana modes is an intriguing property of the driven system. In our case, the dissipation makes it even richer.

\bibliographystyle{apsrev4-2mod}
\bibliography{bibfile.bib}

\end{document}